\documentclass[conference, 10pt]{IEEEtran}
\IEEEoverridecommandlockouts
\usepackage{amsmath}
\usepackage{cite}
\usepackage{amsmath,amssymb,amsfonts}
\usepackage[ruled,linesnumbered,lined]{algorithm2e}
\usepackage{graphicx}
\usepackage{textcomp}
\usepackage{xcolor}
\usepackage{bm}
\usepackage{booktabs}

\usepackage{amsthm,amssymb,amsfonts}

\SetKwInput{Input}{input}
\SetKwInput{Output}{output}

\usepackage{array}
\newcolumntype{L}[1]{>{\raggedright\let\newline\\\arraybackslash\hspace{0pt}}m{#1}}
\newcolumntype{C}[1]{>{\centering\let\newline\\\arraybackslash\hspace{0pt}}m{#1}}
\newcolumntype{R}[1]{>{\raggedleft\let\newline\\\arraybackslash\hspace{0pt}}m{#1}}

\def\BibTeX{{\rm B\kern-.05em{\sc i\kern-.025em b}\kern-.08em
		T\kern-.1667em\lower.7ex\hbox{E}\kern-.125emX}}

\begin{document}
	% \title{An Efficient Waveform Optimization Algorithm for MIMO Radar}
 % \title{A Linear-Proximal Method for MIMO Radar  Waveform Optimization}
 \title{Cramér-Rao Bound Based Waveform Optimization for MIMO Radar:  An Efficient Linear-Proximal Method}
 \author{
		\IEEEauthorblockN{Xiaohua Zhou$^{\star}$,  Xu Du$^{\dagger}$,
			%Kexin Chen$^{\star}$,
			Yijie Mao$^{\star}$ }
		\IEEEauthorblockA{
			$^{\star}$School of Information Science and Technology, ShanghaiTech University, Shanghai 201210, China \\
			$^{\dagger}$Institute of Mathematics HNAS, Henan Academy of	Science, Zhengzhou 450000, China \\
			%		$^{\ddagger }$Department of Electrical and Electronic Engineering,	Imperial College London, United Kingdom\\
			Email:\{zhouxh3, 
			%	chenkx2023, 
			maoyj\}@shanghaitech.edu.cn, duxu@hnas.ac.cn
		}
  \thanks{This work has been supported in part by the National Nature Science Foundation of China under Grant 62201347; in part by Shanghai Sailing Program under Grant 22YF1428400; and in part High-level Talent Research Start-up Project
Funding of Henan Academy of Sciences (Project NO.241819034). \textit{(Corresponding author: Yijie Mao)}}
  }
	% \author{Xiaohua Zhou, \textit{Student Member, IEEE}, Xu Du, \textit{Member, IEEE}, Yijie Mao, \textit{Member, IEEE}
	% 	\thanks{%This work has been supported in part by the National Nature Science Foundation of China under Grant 62201347; and in part by Shanghai Sailing Program under Grant 22YF1428400.
	% 		\par X. Zhou and Y. Mao are with the School of Information Science and Technology, ShanghaiTech University, Shanghai 201210, China (e-mail:
	% 		{zhouxh3, maoyj}@shanghaitech.edu.cn).
	% 		\par X. Du is with Institute of Mathematics HNAS, Henan Academy of	Science, Zhengzhou 450000, China (e-mail: duxu@hnas.ac.cn).
	% 	}

	% }
	
	%\markboth{Journal of \LaTeX\ Class Files,~Vol.~18, No.~9, September~2020}
	%{How to Use the IEEEtran \LaTeX \ Templates}
	
	\maketitle
	
	\begin{abstract}
		 This paper focuses on radar waveform optimization for minimizing the Cramér-Rao bound (CRB) in a multiple-input multiple-output (MIMO) radar system. In contrast to conventional approaches relying on semi-definite programming (SDP) and optimization toolboxes like CVX, we introduce a pioneering and efficient waveform optimization approach in this paper. Our proposed algorithm first applies sequential linear approximation to transform the original CRB-based problem with the transmit power constraint into a sequence of convex subproblems. By introducing a proximal term and further leveraging the Karush-Kuhn-Tucker (KKT) conditions, we derive the optimal closed-form solution for each subproblem. The convergence of the proposed algorithm is then proved rigorously. Numerical results demonstrate that the proposed approach significantly reduces computational complexity — at least two orders of magnitude lower than the baseline algorithms while maintaining the same radar sensing accuracy.

	\end{abstract}  
	

	\section{Introduction}
	Nowadays, positioning and sensing have received considerable attention as key enablers for a series of emerging applications such as augmented reality, virtual reality, the internet of things, autonomous driving, and space-air-ground integrated networks, which simultaneously require high data rate communications and accurate positioning and sensing \cite{BANAFAA2023245}. In this context, radar plays a crucial role by offering advanced capabilities for accurate target detection, localization, and tracking. Multiple-input multiple-output (MIMO) radar, in particular, is favored for its superior performance compared to traditional radar systems.
 %MIMO radar plays a crucial role in this context by offering advanced capabilities for precise target detection, localization, and tracking. The spatial diversity and high resolution provided by MIMO radar systems enhance the accuracy and reliability of positioning and sensing applications. As a result, research on MIMO radar has gained significant momentum, aiming to leverage its potential for improving performance in these emerging applications. The integration of MIMO radar with positioning and sensing technologies is pivotal in meeting the stringent requirements of modern, data-intensive environments, thus driving innovation and advancements in multiple sectors.
  %Multiple-input multiple-output (MIMO) radar is favored for its superior capabilities in target detection, localization, and tracking compared to traditional radar systems. 
  It is highly valuable in applications like defense, surveillance, and remote sensing \cite{9034082, 6649991}. The study of waveform design in the context of MIMO radar has been extensively explored with the aim of improving estimation accuracy \cite{10138058}. 
 Radar waveform design can be classified into the following four categories \cite{9034082}: 1) design waveforms based on information theoretic measures, minimum mean-square error estimation (MMSE), or Cramér-Rao bound (CRB), 2) design waveforms with good auto/cross-correlation properties, 3) design desired transmit beam patterns, and 4) jointly design the transmit waveforms and receive filters for enhancing target detection performance. This paper primarily focuses on the first category, specifically the CRB-based waveform design, which has been widely applied in integrated sensing and communication (ISAC) \cite{9832622, liu2022integrated}, localization \cite{hua2023near}, and underwater acoustic sensor networks (UASNs) \cite{xiong2020layout}, etc.

	CRB, a lower bound on the variance of unbiased estimators, is equivalent to the inverse of the Fisher information matrix (FIM). Since optimizing a matrix as an objective function is impossible, several scalar mapping of FIM has been studied in existing works \cite{TSP2008, pukelsheim2006optimal}, and \cite[Chapter 7.5.2]{boyd2004convex} including: a) the trace of the inverse of FIM, b) the maximum eigenvalue of the inverse of FIM, c) the determinant of  the inverse of FIM, d) the maximal diagonal element of the inverse of FIM. Readers are referred to \cite{pukelsheim2006optimal} for a detailed summary for all mapping criteria.
	For solving the optimization problems with the aforementioned scalar mappings of FIM, several optimization algorithms have been developed \cite{9832622,TSP2008,LongWCNC9771644}. 
These algorithms all share the common approach of approximating the original problem into a semi-definite programming (SDP) by lifting the beamforming vector variabled into positive semi-definite matrix variabled \cite{boyd1997semidefinite,vandenberghe1996semidefinite,ben2021lectures} and relaxing the rank-one constraint. After solving the SDP based on the solvers in optimization toolboxes such as CVX \cite{grant2009cvx},  post processing is carried out to recover the \emph{vector space precoder} based on eigenvalue decomposition (EVD) \cite{9832622, LongWCNC9771644} or Gaussian randomization \cite{luo2010semidefinite, 9652071}.
These approaches, also known as semi-definite relaxation (SDR) \cite{vandenberghe1996semidefinite,freund2004introduction,qi2020semidefinite,mehanna2013joint}, come with significant drawbacks that must be acknowledged. First, employing standard solvers in optimization toolboxes to solve SDP often results in high computational complexity, making it impractical for practical applications. Second, the post processing methods based on EVD or Gaussian randomization are susceptible to numerical instability, potentially resulting in the failure to meet the rank-one constraint. They are also lack of theoretical guarantees to ensure the optimality conditions. 
Third, there are relatively few studies on efficiently solving SDPs, especially for non-convex SDPs. %Note that,  although \cite{jarre2000interior} proposed a matrix variables based \emph{interior method} for  SDPs, its high computational complexity prevents the use of this method.
Therefore, there is an urgent need for efficient algorithms that can directly derive the solution of the beamforming vectors without relying on SDR and optimization toolboxes. Unfortunately, to the best of our knowledge, such an approach has yet to be uncovered in existing literature.

%Semidefinite programming (SDP) is a widely used form of optimization problem, which has corresponding conversion methods to other types of optimization problems, such as linear programming (LP), quadratic programming (QP), and quadratically constrained quadratic programming (QCQP)\cite{vandenberghe1996semidefinite,wen2010alternating,ben2021lectures}. Since the theoretical research on convex SDP is relatively well-developed, many non-convex problems are also approximated and solved as convex SDP \cite{boyd1997semidefinite,luo2010semidefinite,yang2023inexact}. Recently, there have been new relaxation approaches, as discussed in  \cite{lee2021multi,huang2022three,martin2023guarantees}. Specifically, in the fields of communication and radar applications, works such as those in \cite{qi2020semidefinite,mehanna2013joint} have become typical examples of SDP problems. However, there are relatively few studies on efficiently solving SDPs, especially non-convex SDPs \cite{jarre2000interior}, which is extremely high complex. Due to the substantial challenge of developing a low-complexity algorithm for solving non-convex SDP problems generally, this paper attempts to efficiently solve a class of special non-convex SDP problems derived from simplified Multiple-input multiple-output (MIMO) radar scenarios \cite{9832622,TSP2008,LongWCNC9771644}, making a step forward in addressing non-convex SDP challenges.

	This paper aims to closing this gap by proposing an efficient waveform optimization algorithm for the CRB-based radar sensing problems. Inspired by our recent work \cite{Du2022}, we propose an iterative algorithm which employs linear approximation techniques for the objective function and the sequential quadratic programming (SQP) theory \cite{jorge2006numerical} for the transmit power constraint. Additionally, we introduce an $L_2$ norm proximal term \cite{lemaire1989proximal,iusem1999augmented} %also referred to as the proximal point algorithm \cite{iusem1999augmented}, 
	to ensure the convexity of the augmented objective function.

	Contributions of this paper are summarized as follows:
	\begin{itemize}
		\item[a)] A novel optimization algorithm is proposed to tackle the CRB-based waveform design problem with a transmit power constraint. Specifically, we propose an iterative algorithm by solving a sequence of approximated convex subproblems. A closed-from solution is obtained for each subproblem, which therefore eliminates the need for optimization toolbox. Moreover, the proposed algorithm derives the beamforming vectors' solution directly without the need for SDR.
		\item[b)] The sub-linear convergence property \cite[Thm 2.1.14]{nesterov2018lectures} of the proposed algorithm is then proved rigorously. 
		\item[c)] Numerical results are conducted to demonstrate that the proposed algorithm significantly reduces CPU time by at least two orders of magnitude compared to traditional methods while maintaining comparable radar sensing performance.
	\end{itemize}

\textit{Organizations}: The rest of the following paper is structured as follows. Section $\text{\uppercase\expandafter{\romannumeral2}}$ presents the system model. Section $\text{\uppercase\expandafter{\romannumeral3}}$ discusses problem formulation and the existing algorithms. In Section $\text{\uppercase\expandafter{\romannumeral4}}$, we introduce the proposed \textit{Linear-Proximal Method} to solve the non-convex SDP problem. Section $\text{\uppercase\expandafter{\romannumeral5}}$ presents the simulation results, followed by conclusion in Section $\text{\uppercase\expandafter{\romannumeral6}}$.

\textit{Notations}: Bold upper and lower case letters denote matrices and column vectors, respectively. $\Re(\cdot)$ and $\Im(\cdot)$ respectively denote the real and imaginary parts of a complex scalar, vector, or matrix. $ (\cdot)^\top $, $ (\cdot)^H $, $ |\cdot| $, $ \| \cdot \| $, $ (\cdot)^{-1} $, and $ \operatorname{tr}(\cdot) $ represent the transpose, Hermitian, absolute value, Euclidean norm, inverse, and trace operators, respectively. 
	
	% \section{Preliminaries}\label{sec:pre}
	% In this section, the basics of the considered MIMO radar model and CRB-based waveform optimization problems are reviewed followed by a summary of the conventional algorithms to solve the formulated waveform optimization problems. 

	\section{System model}\label{sec: Wireless Communication Model}
We consider a monostatic radar system equipped with $ N_t $ transmit antennas and $ N_r $ receive antennas, aiming to detect the position of a single target\footnote{For simplified illustration, we consider a single target in this work. However, the proposed algorithm can be easily extended to a multi-target scenario.}. There are $L$ transmission and radar pulse blocks in one coherent processing interval (CPI) indexed by $\mathcal{L}=\{1, \ldots, L\}$. The overall  transmit signal from the MIMO radar is $ \mathbf{X} = [\mathbf{x}[1], \mathbf{x}[2], \cdots, \mathbf{x}[L]] \in \mathbb{C}^{N_t \times L} $. In each transmission block $l$, this MIMO radar receives the following reflected echo from the target:	
  \begin{equation}
  	\mathbf{y}_r[l] = \beta e^{j2\pi \mathcal{F}_DlT} \mathbf{a}_r(\theta) \mathbf{a}_t^H(\theta) \mathbf{x}[l] + \mathbf{z}_r[l],
  \end{equation}
where $\beta$ represents the reflection
coefficient, measuring the portion of the incident electromagnetic wave reflected back to the MIMO radar. $\mathcal{F}_D $ represents the Doppler frequency. \vspace{2mm}It is calculated by $ \mathcal{F}_D = \frac{2vf_c}{c} $, where $ v $ is the velocity of the target, $ f_c $ is the carrier frequency, and $ c $ denotes the speed of light. $ \mathbf{z}_r[l] \in \mathbb{C}^{N_r \times 1} $ is an additive white Gaussian noise vector with each element having a zero mean and a variance of $\sigma_r^2$. $ \mathbf{a}_t(\theta) \in \mathbb{C}^{N_t \times 1} $ and $ \mathbf{a}_r(\theta) \in \mathbb{C}^{N_r \times 1} $ respectively represent the steering vectors
for the transmit and receive antenna array, which are defined as:
\begin{equation*}
	\mathbf{a}_t(\theta)=[1, e^{j\pi \sin (\theta)}, \cdots,  e^{j\pi (N_t - 1) \sin (\theta) }], 
 \end{equation*}
 \begin{equation*}
	\mathbf{a}_r(\theta)=[1, e^{j\pi \sin (\theta)}, \cdots,  e^{j\pi (N_r - 1) \sin (\theta) }], 
\end{equation*}
where $ \theta $ denotes the angle of departure and arrival for the line of sight (LoS) path of the monostatic radar.

 The covariance of the transmit signal is denoted as:
\begin{equation}\label{eq: rank1}
	\mathbf{P} \overset{\text{def}}{=} \frac{1}{L}\sum_{l=1}^L  \mathbf{x}[l]\mathbf{x}[l]^H = \mathbf{p}\mathbf{p}^H,
\end{equation}
where $\mathbf{p} \in \mathbb{C}^{N_t \times 1}$ is the beamforming vector of the transmit signal.

Define the vector for estimation parameters as $ \bm \xi =[\theta, \beta^\Re, \beta^\Im, \mathcal{F}_D]^T $, where $\theta$ is the direction of departure (DoD) and direction of arrival (DoA) of the target, $\beta^\Re$ denotes the real part of the complex coefficient, $\beta^\Im$ represents the imaginary part of the complex coefficient, and $\mathcal{F}_D$ is the doppler frequency. The FIM matrix used to estimate all parameters in $\bm \xi$ is given as \cite{9832622}:
\begin{equation}\label{eq: FIM}
	\mathbf{F}(\mathbf{p}) =  \begin{bmatrix}
		F_{11} & F_{12} & F_{13} & F_{14} \\
		F_{21} & F_{22} & F_{23} & F_{24} \\
		F_{31} & F_{32} & F_{33} & F_{34} \\
		F_{41} & F_{42} & F_{43} & F_{44}
	\end{bmatrix},
\end{equation}
where the elements of FIM are evaluated as
\begin{equation}\label{eq:F-elment}
	F_{ij}=\frac{2}{\sigma_r^2}\Re\left\{\sum_{l=1}^L\frac{\partial \bm \mu^\top[l]}{\partial \bm \xi_i} \frac{\partial \bm \mu[l]}{\partial \bm \xi_j}\right\}, \forall i,j = 1, \cdots, 4, 
\end{equation}
where $ \bm \xi_i$, $\bm \xi_j$ are the $i$th and the $j$th elements of $\bm \xi$, and $ \bm \mu[l]= \mathbf{y}_r[l]-\mathbf{z}_r[l] $.
	
%Due to matrix optimization complexities, several approaches have explored scalar mappings of CRB-based waveform optimization problems, describing CRB as $\mathbf{F}^{-1}$. 

\section{Problem Formulation and Existing Algorithms}\label{sec: Tradition Methods for OED}

Based on the FIM matrix defined in (\ref{eq: FIM}), we then obtain the CRB metric as $\mathbf{F}^{-1}$. There are two classical transmit beamforming optimization problems based on CRB: minimizing the trace of inverse of FIM \cite{10138058} and minimizing the largest eigenvalue of the inverse of FIM \cite{9832622}, both subject to the transmit power constraint.

The optimization problem based on minimizing the trace of inverse of FIM is formulated as:
	%This paper focuses on the optimization problem based on the former problem, which is formulated as: 
	\begin{equation}\label{eq:non-convex original pro}
		\begin{split}
			&\min_{\mathbf{p}} \quad \operatorname{tr}(\mathbf{F}^{-1})\\
			&\operatorname{s.t.} \quad \operatorname{tr} \left( \mathbf{p}\mathbf{p}^H \right) \leq P_t. 
		\end{split}
	\end{equation}

	To circumvent the need for inversion in Problem \eqref{eq:non-convex original pro}, existing works typically utilize the Schur complement to reformulate the problem as \cite{LongWCNC9771644}: 
\begin{equation}\label{eq: Baseline Algorithm 1}
		\begin{split}
			&\min_{\mathbf{P}, t_1, t_2, t_3, t_4} \quad \sum_{i=1}^4 t_i \\
			&\operatorname{s.t.} \begin{bmatrix}
				\mathbf{F} & \mathbf{e}_i \\
				\mathbf{e}_i^T & t_i 
			\end{bmatrix} \succeq 0, i \in \{1,2,3,4\}, \\
			& \operatorname{tr} \left( \mathbf{P} \right) \leq P_t, \\
			& \operatorname{rank}(\mathbf{P}) = 1,
		\end{split}
	\end{equation}
	where $t_1, \cdots, t_4$ are auxiliary variables, and $\mathbf{e}_i$ is the $i$-column of a $4 \times 4$ identity matrix.

The optimization problem based on minimizing the largest eigenvalue of FIM is given as: \begin{equation}\label{eq: E-criterion}
		\begin{split}
			&\min_{\mathbf{p}} \quad \sigma_{\max} (\mathbf{F}^{-1}) \\
			&\operatorname{s.t.} \quad \operatorname{tr} \left( \mathbf{p}\mathbf{p}^H \right) \leq P_t.
		\end{split}
	\end{equation}

By introducing an auxiliary variable $t$, Problem \eqref{eq: E-criterion} can be equivalently reformulated as \cite{vandenberghe1996semidefinite}:
	\begin{equation}\label{eq: Baseline Algorithm 2}
		\begin{split}
			&\max_{\mathbf{P},t} \quad t \\
			\operatorname{s.t.}& \quad \mathbf{F} \succeq t\mathbf{I}, \\
			& \operatorname{tr} \left( \mathbf{P} \right) \leq P_t,\\
			& \operatorname{rank}(\mathbf{P}) = 1.
		\end{split}
	\end{equation}
 
Since Problem \eqref{eq:non-convex original pro} minimizes the sum of all eigenvalues of the inverse of  FIM, while Problem \eqref{eq: E-criterion} only minimizes the maximum eigenvalue of the inverse of FIM, the CRB performance obtained by solving Problem \eqref{eq:non-convex original pro} is always better than the one obtained from solving Problem \eqref{eq: E-criterion}. Therefore, Problem \eqref{eq:non-convex original pro} is preferred in most literature \cite{TSP2008,luzhou07,LongWCNC9771644}. This paper focuses on addressing Problem \eqref{eq:non-convex original pro} efficiently.

%Note that, in most literature \cite{TSP2008,luzhou07,LongWCNC9771644}, we employ Problem \eqref{eq:non-convex original pro} to measure CRB. This is because even both Problem \eqref{eq: E-criterion} and \eqref{eq:non-convex original pro} are scalar mapping of FIM, \eqref{eq:non-convex original pro} minimizes the sum of all eigenvalues of FIM, such that the performance is generally better compare with \eqref{eq: E-criterion} in practice. Therefore, this paper focuses on the optimization problem based on the former problem.}}

For Problem \eqref{eq:non-convex original pro}, the reformulated problem \eqref{eq: Baseline Algorithm 1} is a convex SDP if removing the rank-one constraint. Therefore, the optimal covariance matrix $\mathbf{P}$ for the relaxed problem without the rank-one constraint can be directly obtained by the standard optimization toolbox like CVX \cite{grant2009cvx}. Such approach is known as SDR.  However, removing the rank-one constraint in \eqref{eq: Baseline Algorithm 1} may lead to a solution $\mathbf{P}$ that is not rank-one. Although the transmit beamforming solution $\mathbf{p}$ can be restored through EVD \cite{9832622} or Gaussian randomization \cite{9652071}, the solution might not be optimal for the original non-convex problem. Furthermore, solving SDP via CVX toolbox essentially employs interior-point methods, which are known for their high computational complexity.

	In this paper, the following strategies are demonstrated as the baselines.
	For later convenience, we abbreviate them as follows: a) \emph{SDP-EVD} for the method of using SDR and EVD to solve Problem \eqref{eq: Baseline Algorithm 1}; b) \emph{SDP-random} for the approach that solves Problem \eqref{eq: Baseline Algorithm 1} by SDR combined with Guassian randomization; c) \emph{Upper-Bound} for the approach which removes the rank-one constraint in \eqref{eq: Baseline Algorithm 1},
and solves the corresponding relaxed convex problem using the standard optimization toolbox CVX. The aforementioned first two baselines are summarized in Algorithm 1. The third baseline is obtained by simply removing step 2 of Algorithm 1.

	{
		\SetNlSty{textbf}{}{:}
		\IncMargin{1em}
		\begin{algorithm}[h]
		%	\small
			\KwIn{$\theta, \beta^\Re, \beta^\Im, \mathcal{F}_D$;}
			
			Update $ \mathbf{P} $ by removing the rank-one constraint in Problem \eqref{eq: Baseline Algorithm 1} and solving the corresponding relaxed problem via CVX toolbox;\\
			Calculate $ \mathbf{p} $ through EVD/Guassian randomization.\\
			%Calculate $\mathcal{F}^{-1}$;
               \KwOut{ $\mathbf{p} $;}
			\caption{Baseline 1-2}
			\label{alg：baseline1}
		\end{algorithm}
	}
	
	\section{Proposed Algorithm}\label{sec: proposed algorithm}
	In this section, to circumvent the reliance on optimization toolboxes and address the issue of matrix inversion in \eqref{eq:non-convex original pro}, we propose a novel algorithm named \emph{Linear-Proximal Method} (LPM), which iteratively solves a sequence of convex subproblems. In the end, we provide the corresponding convergence analysis.
	
	\subsection{Linear-Proximal Method}

In order to deal with the symbolic inverse of  FIM in the objective function of \eqref{eq:non-convex original pro} while avoiding the matrix inequalities and matrix variables as in \eqref{eq: Baseline Algorithm 1}, inspired from \cite[Equation 9]{Du2022}, we adopt a tight linearized approximation of $\operatorname{tr}(\mathbf{F}^{-1})$ to approximate Problem \eqref{eq:non-convex original pro} into a sequence of subproblems. The subproblem in the $(k+1)$-th iteration is:
	\begin{equation}\label{eq: linear-app1}
		\begin{split}
			&\min_{\mathbf{p}} \quad \operatorname{tr}\left(  \left(\mathbf{F}^k\right)^{-1} \right)- \operatorname{tr}\left( \left(\mathbf{F}^k\right)^{-1}  \mathbf{F} \left(\mathbf{F}^k\right)^{-1} \right)\\
			&\operatorname{s.t.} \quad \operatorname{tr} \left( \mathbf{p}\mathbf{p}^H \right) \leq P_t,
		\end{split}
	\end{equation}
	where $ \mathbf{F}^k $ represents the numerical value matrix of $ \mathbf{F} $ derived based on the solution $\mathbf{p}^k$ in the $(k)$-th iteration. 
 
Following the standard SQP theory \cite[Chapter 18]{jorge2006numerical},  linearization is applied to the transmit power constraint of \eqref{eq: linear-app1},  which approximates the quadratic term $\operatorname{tr}(\mathbf{p}\mathbf{p}^H)$ to $\operatorname{tr} ( \mathbf{p}^k ( \mathbf{p}^k )^H ) + 2 ( \mathbf{p}^k )^H ( \mathbf{p} - \mathbf{p}^k )$ with the solution of $\mathbf{p}^k$ obtained from the ($k$)-th iteration. This linearization also adds the term $  \lambda^k \mathbf{p}^H\mathbf{p} $ to the objective function guaranteeing that the KKT conditions are satisfied, where $\lambda^k$ denotes the value of the dual variable $\lambda$ introduced by the transmit power constraint in Problem \eqref{eq: linear-app1} during the $ (k) $-th iteration. Problem \eqref{eq: linear-app1} is then equivalently transformed to:
	\begin{equation}\label{eq: con-approx}
		\begin{split}
			&\min_{\mathbf{p}} \quad \lambda^k \mathbf{p}^H\mathbf{p} - \operatorname{tr}\left( \left(\mathbf{F}^k\right)^{-1}  \mathbf{F} \left(\mathbf{F}^k\right)^{-1} \right) \\
			&\operatorname{s.t.} \quad \operatorname{tr} \left( \mathbf{p}^k \left( \mathbf{p}^k \right)^H \right) + 2 \left( \mathbf{p}^k \right)^H \left( \mathbf{p} - \mathbf{p}^k \right) = P_t|\lambda.
		\end{split} 
	\end{equation}
 
	Unfortunately, 
 the objective function of Problem \eqref{eq: con-approx} remains non-convex. %objective function of Problem \eqref{eq: con-approx} remains non-convex due to the complex FIM $\mathbf{F}$ defined in \eqref{eq: FIM}. 
 To address the issue, we introduce an $ L_2 $-norm proximal term in the objective function. Moreover, the term $\operatorname{tr}( (\mathbf{F}^k)^{-1}  \mathbf{F} (\mathbf{F}^k)^{-1} )$ is also equivalently transformed to $\sum_{i=1}^4 \sum_{j=1}^4 a_{ij}F_{ji}$, where $a_{ij}, \forall i,j \in \{1, \cdots, 4\}$ denote the elements of $ \left(\mathbf{F}^k\right)^{-1} \left(\mathbf{F}^k\right)^{-1} $, and $F_{ji}, \forall i,j=1,...,4 $ are defined in \eqref{eq:F-elment}.
%i.e. $\left(\mathbf{F}^k\right)^{-1} \left(\mathbf{F}^k\right)^{-1} = \begin{bmatrix}	a_{11} & a_{12} &a_{13} & a_{14} \\	a_{21} & a_{22} &a_{23} & a_{24} \\	a_{31} & a_{32} &a_{33} & a_{34} \\	a_{41} & a_{42} &a_{43} & a_{44} \end{bmatrix} $.
The resulting problem becomes:
\begin{equation}\label{eq: obj-approx}
	\begin{split}
		\min_{\mathbf{p}} \quad
		&- \sum_{i=1}^4 \sum_{j=1}^4 a_{ij}F_{ji} 
		+ \lambda^k \mathbf{p}^H\mathbf{p} + \frac{\rho}{2} \|\mathbf{p} - {\mathbf{p}^k} \|^2 \\ 
		\operatorname{s.t.} \quad &\operatorname{tr} \left( \mathbf{p}^k\left( \mathbf{p}^k \right)^H \right) + 2 \left( \mathbf{p}^k \right)^H \left( \mathbf{p} - \mathbf{p}^k \right) = P_t\;|\;\lambda,
	\end{split} 
\end{equation}
where $\rho>0$ is a sufficient penalty coefficient. With a sufficient large $\rho$, the proximal term $\frac{\rho}{2} \|\mathbf{p} - {\mathbf{p}^k} \|^2$ ensures the convexity of the objective function, thereby facilitating stable numerical convergence of the algorithm.

	Note that, the Lagrange function of the convex Problem \eqref{eq: obj-approx} is obtained as follows:
\begin{equation*}\label{eq:Lagrangefun}
	\begin{split}
		\mathcal{L}(\mathbf{p}, \lambda) = 
		&- \sum_{i=1}^4 \sum_{j=1}^4 a_{ij}F_{ji} + \lambda^k \mathbf{p}^H\mathbf{p} + \frac{\rho}{2} \|\mathbf{p} - {\mathbf{p}^k} \|^2 \\ 
	&+\lambda \left[ \operatorname{tr}\left(\mathbf{p}^k\left( \mathbf{p}^k \right)^H\right)  + 2 \left( \mathbf{p}^k \right)^H \left( \mathbf{p} - \mathbf{p}^k \right) - P_t\right],
	\end{split} 
\end{equation*}
while the corresponding Karush-Kuhn-Tucker (KKT) stationary condition is expressed as:	
	\begin{equation}\label{eq: original-KKT-prox}
		\left\{
		\begin{aligned}
			\frac{\partial \mathcal{L}}{\partial \mathbf{p}} =& \mathbf{Q}^k \mathbf{p} + 2 \lambda \mathbf{p}^k - \rho \mathbf{p}^k = \bm 0 \\
			\frac{\partial \mathcal{L}}{\partial \lambda} =& \operatorname{tr} \left( \mathbf{p}^k \left( \mathbf{p}^k \right)^H \right) + 2 \left( \mathbf{p}^k \right)^H \left( \mathbf{p} - \mathbf{p}^k \right) - P_t = 0, 
		\end{aligned} \right.
	\end{equation}
	where $\mathbf{Q}^k = - \frac{ 1 }{\sigma_r^2} \bm{\Theta} + 2 \lambda^k \mathbf{I} + \rho \mathbf{I}$, $\mathbf I$ denotes the identity matrix, and $\bm{\Theta}$ is given as
% 	\begin{equation}\label{eq: Q}
% 	\begin{split}
% 		\mathbf{Q}^k = - \frac{ 1 }{\sigma_r^2}   \varTheta + 2 \lambda^k \mathbf{I} + \rho \mathbf{I},
% 	\end{split}
% \end{equation} 
\begin{equation*}
    \begin{split}
       \bm{\Theta} =& b \left(\frac{\partial \mathbf{A}(\theta)}{\partial \theta}\right)^H \frac{\partial \mathbf{A}(\theta)}{\partial \theta}  + c \mathbf{A}^H(\theta) \frac{\partial \mathbf{A}(\theta)}{\partial \theta}\\
        &+ d \left(\frac{\partial \mathbf{A}(\theta)}{\partial \theta }\right)^H  \mathbf{A}(\theta)    + e \mathbf{A}^H(\theta) \mathbf{A}(\theta).
        \end{split}
\end{equation*}
Here, $\mathbf{A}(\theta)=\mathbf{a}_r(\theta) \mathbf{a}_t^H(\theta)$. $b, c, d, e$ in $\bm \Theta$ are given as:
\begin{equation*}
\begin{split}
        b=&2L|\beta|^2 a_{11},\\
        c=&L \beta a_{21} - L \beta a_{31}j - \pi (L+1) |\beta|^2a_{41}j + L \beta a_{12}\\
        &- L \beta a_{13}j - \pi(L+1)|\beta|^2 a_{14}j,\\
        d=& L \beta^* a_{21}+ L \beta^* a_{31}j + \pi (L+1) |\beta|^2a_{41}j + L \beta^* a_{12}\\
        &+ L \beta^* a_{13}j+\pi(L+1)|\beta|^2 a_{14}j,\\
        e=&2La_{22}+\pi(L+1)(\beta-\beta^*)a_{42}j+\pi(L+1)(\beta+\beta^*)a_{43}\\
        &+\pi(L+1)(\beta-\beta^*)a_{24}j+\pi(L+1)(\beta+\beta^*)a_{34}\\
        &+2La_{33} + \left( \frac{4\pi^2}{L} \right) \frac{(L+1)(2L+1)}{3}|\beta|^2a_{44}.
    \end{split}
\end{equation*}

%	 Here, we introduce a remark to specify the value of $\rho$.
%	 \begin{remark}
%	 	From Equation \eqref{eq: Q}, the second order information of the objective of Problem \eqref{eq: con-approx} $ \tilde{\mathbf{Q}}^k \overset{\text{def}}{=} \mathbf{Q}^k- \rho \mathbf{I} $,
%	 	where $\mathbf{Q}^k\succ 0$ (second order sufficient condition, SOSC \cite{jorge2006numerical}).
%	 	Therefore, $\rho$ has at least a lower bound such that
%	 $ \rho\geq \mathcal E_{\text{min}}\left( \tilde {\mathbf{Q}}^k\right) $
%	 	with  $\mathcal E_{\text{min}}$ denoted by the corresponding smallest eigenvalue operator.
%	 \end{remark}
	By solving Equation \eqref{eq: original-KKT-prox}, the optimal dual and primal variables are obtained as:
	\begin{subequations}\label{eq:variable update}
		\begin{align}
		%	\lambda^{k+1} =\;& \frac{1}{4}\left( \left(\mathbf{p}^k\right)^H (\mathbf{Q}^k)^{-1} \mathbf{p}^k \right)^{-1} \left( P_t - \text{tr}\left( \mathbf{p}^k \left( \mathbf{p}^k \right)^H \right) \right) \nonumber\\
		%	+& \frac{1}{2}\left( \left( \mathbf{p}^k \right)^H (\mathbf{Q}^k)^{-1} \mathbf{p}^k \right)^{-1}  \left( \mathbf{p}^k \right)^H \mathbf{p}^k + \frac{\rho}{2} \mathbf{I},\nonumber\\
  \lambda^{\star} =\;& \frac{\rho}{2}-\frac{P_t+\left( \mathbf{p}^k \right)^H \mathbf{p}^k}{4\left( \mathbf{p}^k \right)^H\left(\mathbf{Q}^k \right)^{-1}\mathbf{p}^k}\label{eq: lambda-update},\\
			\mathbf{p}^{\star} =\;& (\rho-2 \lambda^{k+1}) (\mathbf{Q}^k)^{-1} \mathbf{p}^k. \label{eq: p-update}
		\end{align}
	\end{subequations}
	We then update $\lambda^{k+1}$ and $\mathbf{p}^{k+1}$ using \eqref{eq:variable update}. The details of the proposed LPM is illustrated in Algorithm \ref{alg3}.  Specifically, the  dual variable $\lambda$ and the  beamforming vector $ \mathbf{p} $ are iteratively updated until the convergence of the objective function $obj^k$ of Problem \eqref{eq: linear-app1}, $\epsilon$ is the convergence tolerance.
\addtolength{\topmargin}{0.03in}

{
	\SetNlSty{textbf}{}{:}
	\IncMargin{1em}
	\begin{algorithm}[h]
		%\small
		\KwIn{$\theta, \beta^\Re, \beta^\Im, \mathcal{F}_D$;}
		\textbf{Initialization: }{Initialize feasible $ \mathbf{p}^{0} $ and $ \mathbf{F}^{0} $, initial iteration index $ k \leftarrow 0 $ and tolerance $ \epsilon > 0 $;}\\
		\Repeat{${|\text{obj}^{k+1} - \text{obj}^{k}| \leq \epsilon} $}
		{
			%Calculate the objective function $\text{obj}^{k}$ in Problem \ref{eq: linear-app1};\\
			Update $ \lambda^{k+1} $ by Equation \eqref{eq: lambda-update};\\
			Update $ \mathbf{p}^{k+1} $ by Equation \eqref{eq: p-update};\\
			Evaluate the objective function $\text{obj}^{k+1}$ in Problem \eqref{eq: linear-app1};
		}
		Set $ \mathbf{p}^\star \leftarrow \mathbf{p}^{k+1}$;\\
		\textbf{return}  $\mathbf{p}^\star $.\\
        \KwOut{  $\mathbf{p}^\star $ ;}
		\caption{Linear-Proximal Method (LPM)}
		\label{alg3}
	\end{algorithm}
}

	\subsection{Convergence Analysis and Computational Complexity}\label{sec: analysis}
	
	\subsubsection{Convergence Analysis}\label{subsec:convergence}
	Next,  we analyze the convergence of Algorithm \ref{alg3}. In the following, we assume that $\mathbf{p}^0$ is initialized in the
	 neighborhood of a local optimal solution $\mathbf{p}^*$. 
	We can then show the sub-linear convergence of Algorithm \ref{alg3} by checking the limit of the following equation:
	\begin{equation}\label{eq: limit}
		\begin{split}
			\frac{\left\| \mathbf{p}^{k+1} -\mathbf{p}^* \right\|}{\left\| {\mathbf{p}}^k  -\mathbf{p}^* \right\|} 
			\overset{\eqref{eq: p-update}}{=}& \frac{\left\| (\rho-2 \lambda^{k+1}) (\mathbf{Q}^{k})^{-1} {\mathbf{p}}^k -\mathbf{p}^* \right\|}{\left\| {\mathbf{p}}^k -\mathbf{p}^* \right\|} .
		\end{split}
	\end{equation}
	As $k$ approaches infinity, the coefficient of $\mathbf{p}^k$ in the numerator  of \eqref{eq: limit} shows:
	\begin{equation}\label{eq: numerator}
		\begin{split}
			&\lim_{k\rightarrow \infty} (\rho-2 \lambda^{k+1}) (\mathbf{Q}^{k})^{-1} \\
			\overset{\eqref{eq: lambda-update}}{=}&\lim_{k\rightarrow \infty}  \frac{1}{2}\left( ({\mathbf{p}}^k)^H (\mathbf{Q}^{k})^{-1} {\mathbf{p}}^k \right)^{-1} \left( P_t + \left( {\mathbf{p}}^k \right)^H  {\mathbf{p}}^k \right) (\mathbf{Q}^{k})^{-1} \\
			=&\lim_{k\rightarrow \infty} \frac{\left( \left( {\mathbf{p}}^k \right)^H  {\mathbf{p}}^k (\mathbf{Q}^{k})^{-1}  \right)}{\left\|\mathbf{p}^k \right\|^2_{(\mathbf{Q}^{k})^{-1}} }
			=\mathbf{I},
		\end{split}
	\end{equation}
where $\|\cdot\|_{\mathbf{M}}$ denotes the Mahalanobis distance with  symmetric positive definite matrix $\mathbf M$ and it simplifies to the  $L_2$ norm when $\mathbf{M}=\mathbf{I}$. The last equality results from the proximity of eigenvalues of matrix $\mathbf Q$ caused by parameter $\rho$ as well as the initial point $\mathbf{p}^0$. Based on \eqref{eq: limit} and \eqref{eq: numerator}, we then obtain the sub-linear convergence property of Algorithm \ref{alg3} by showing: 
	\begin{equation}\label{eq: sublinear}
		\lim_{k\rightarrow \infty}  \frac{\left\| \mathbf{p}^{k+1} -\mathbf{p}^*\right\|}{\left\| \mathbf{p}^k -\mathbf{p}^*\right\| } = 1.
	\end{equation}
	%As a complement, a numerical convergence result that corresponds to the above analysis is illustrated in Fig. \ref{fig:Iteration}. As algorithm \ref{alg3} iterates, the value of \eqref{eq: limit} gets closer and closer to 1.
	% \begin{figure}[h]
	% 	\centering
	% 	\includegraphics[width=0.4\textwidth]{figure/Iteration-Convergence.eps}
	% 	\caption{Convergence behavior of the proposed algorithm, where $P_t = 10$ dBm, $N_t = 8$ and $\epsilon = 10^{-7}$.}
	% 	\label{fig:Iteration}
	% \end{figure}	
	
	\subsubsection{Computational Complexity}
	The complexities of both Algorithm 1 and Algorithm 2 are summarized here. In Algorithm \ref{alg：baseline1}, the SDP problem is solved using interior point methods by CVX, incurring a computational complexity of $\mathcal{O}(\log(N_t^7)$ \cite{LongWCNC9771644, 1634819}. In Algorithm \ref{alg3}, each iteration for obtaining the closed-form solution $\mathbf{p}$ requires a complexity of $\mathcal{O}(N_t^3)$. With the convergence accuracy $ \epsilon $, the overall complexity of Algorithm \ref{alg3} is $\mathcal{O}(\log(\epsilon^{-1}) N_t^3)$.

	\section{Numerical Result}\label{sec: NR}
	In this section, we illustrate the numerical results of the proposed Linear-Proximal algorithm. Three baseline algorithms are considered, as specified in Section \ref{sec: Tradition Methods for OED}. The implementation of all the experiments relies on Windows 11 professional version, MATLAB 2022a and an Intel(R) Core(TM) i7-10700 CPU @ 2.90GHz processor. To mitigate the influence of chance and computer errors, all experimental results are averaged over 100 random trials.

Unless otherwise specified, we consider a MIMO radar with 9 receiver antennas ($ N_r = 9 $), the radar noise power is $ \sigma_r^2 = 0 $ dBm. The true target parameters are $ \theta = 45^{\circ} $ with a velocity of $v=8$ m/s. A total number of $1024$ transmit symbols is considered within one CPI. The proposed algorithm is configured with a penalty coefficient of $ \rho = 5 $ and a convergence tolerance of $\epsilon = 10^{-5}$. The proposed algorithm initializes the transmit beamforming vector based on the direction of the transmission steering vector \cite{765552}. Following \cite{9832622}, the radar SNR is defined as $\frac{|\beta|^2P_t}{\sigma_r^2}=10$ dB, here $|\beta|=1$.

	\begin{figure}[t]
		\centering
		\includegraphics[width=0.45\textwidth]{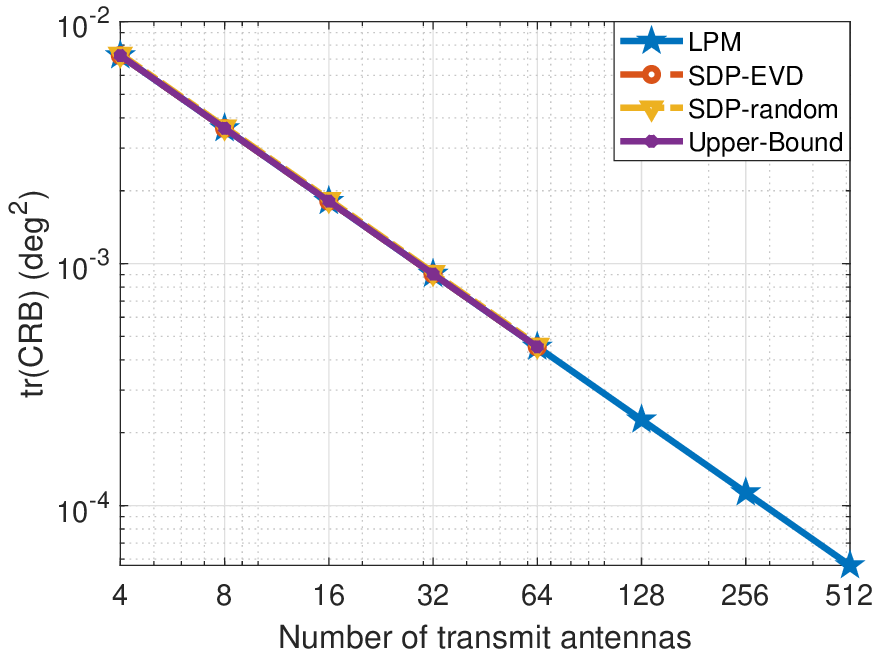}
		\caption{CRB versus the number of transmit antennas when $P_t = 10$ dBm.}
		\label{fig:CRB-Nt}
	\end{figure}

	\begin{figure}[t]
		\centering
		\includegraphics[width=0.45\textwidth]{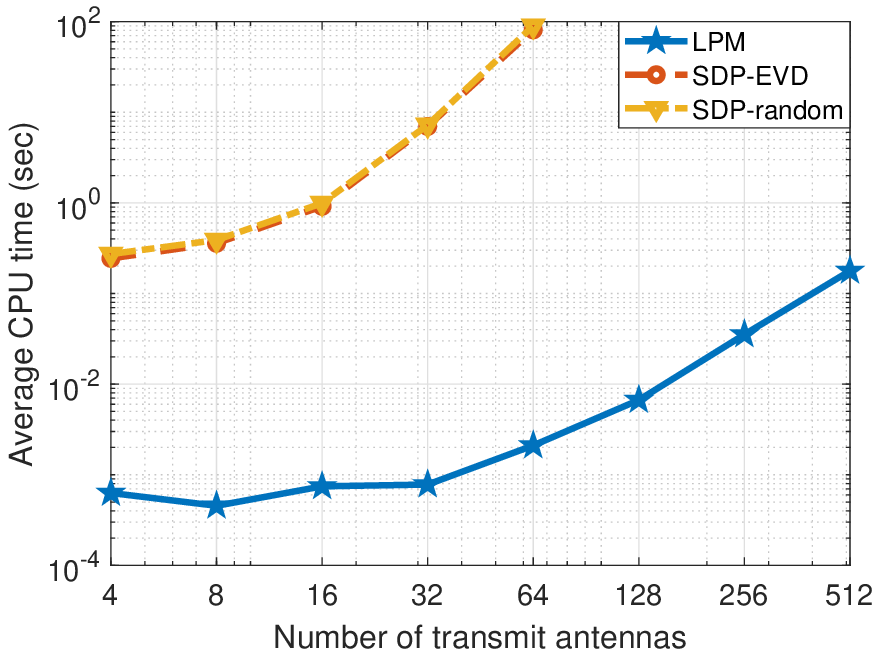}
		\caption{CPU time versus the number of transmit antennas when $P_t = 10$ dBm.}
		\label{fig:T-Nt}
	\end{figure}

The detection performance of the proposed algorithm and the baseline algorithms is depicted in Fig. \ref{fig:CRB-Nt}. 
%Since the relevant value of the global optimal solution in this field is not found, we only show the comparison of the objective function of LPM with the baseline algorithm. 
Due to the complexity of CVX-based algorithms, we only illustrate their results in the range of 4 to 64 transmit antennas. Note that the Upper-Bound does not provide a solution for the transmit beamforming vector $\mathbf{p}$; rather, it only yields the optimal covariance matrix $\mathbf P=\mathbf p \mathbf p^H$. Consequently, such method cannot directly address Problem \eqref{eq:non-convex original pro}. We only consider it as the performance upper bound. From Fig. \ref{fig:CRB-Nt}, we observe that as the number of transmit antennas increases, the trace of the CRB for target estimation decreases for all algoirthms. It is evident that the proposed LPM achieves the same performance as all other benchmark algorithms, including the Upper-Bound algorithm.

 	\begin{figure}[t]
		\centering
		\includegraphics[width=0.45\textwidth]{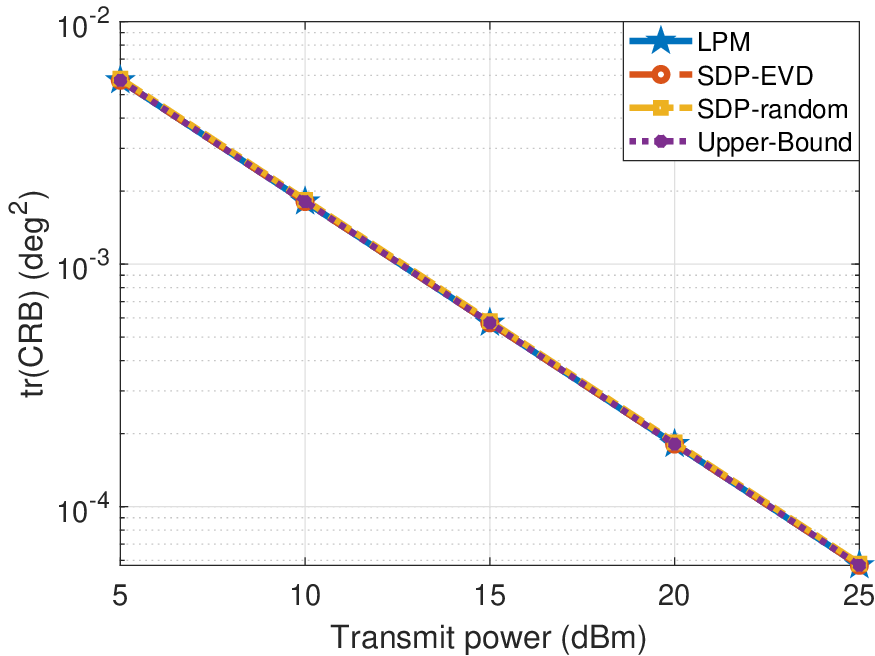}
		\caption{CRB versus the transmit power when $N_t = 16$.}
		\label{fig:CRB-Pt}
	\end{figure}

  	\begin{figure}[t]
		\centering
		\includegraphics[width=0.45\textwidth]{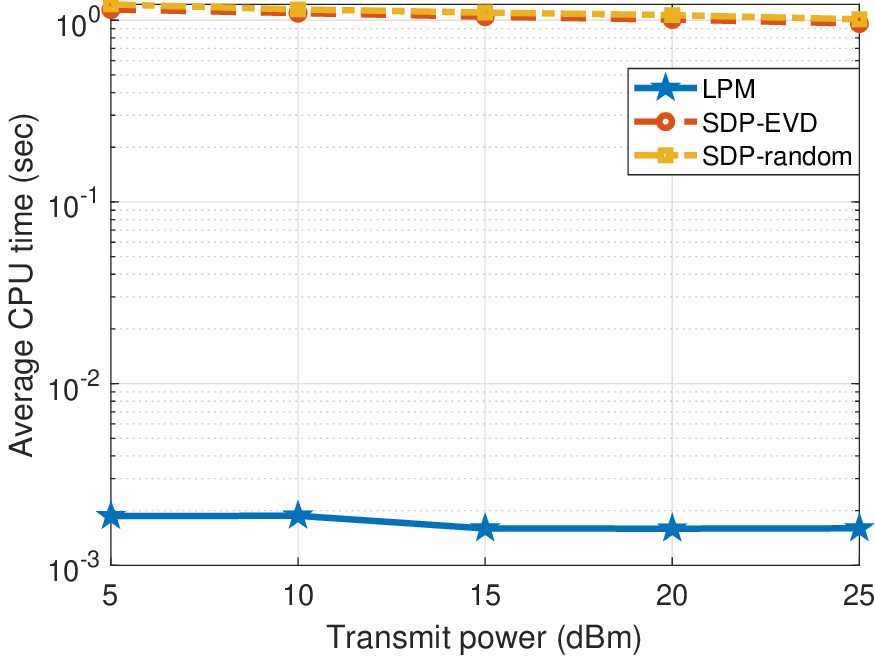}
		\caption{CPU time versus the transmit power when $N_t = 16$.}
		\label{fig:T-Pt}
	\end{figure}
 
 \par In Fig. \ref{fig:T-Nt}, we illustrate the average CPU time consumption for all algorithms as the number of antennas increases. It should be noted that although Section \ref{sec: analysis} indicates that the convergence rate of LPM is only sub-linear, the proposed LPM reduces the average CPU time by at least two orders of magnitude compared to the benchmark algorithms. Since the Upper-Bound algorithm cannot solve Problem \eqref{eq:non-convex original pro}, its CPU time is meaningless, and thus omitted from Fig. \ref{fig:T-Nt}.
 
 \par Fig. \ref{fig:CRB-Pt} illustrates the CRB performance for all algorithms, covering a range of the transmit power from 5 dBm to 25 dBm. As the transmit power increases, the CRB value decreases for all algorithms, indicating an enhancement of the target detection accuracy. Additionally, the performance of our proposed algorithm aligns with that of the baseline algorithms.

 \par In Fig. \ref{fig:T-Pt}, we present the average CPU time consumption for all algorithms as the transmit power increases. It is evident that our proposed algorithm significantly reduces the average CPU time compared to the baseline algorithms.

	\section{Conclusion $\&$ Outlook}\label{sec: Conclusion}
	In this paper, for designing waveform in MIMO radar systems, a novel low-complexity algorithm to solve non-convex SDP problem is proposed based on linear approximation and proximal term. As a theoretical contribution, we also prove the sub-linear convergence property of the proposed algorithms. Numerical results demonstrate that the proposed algorithm significantly reduces the CPU time while maintaining comparable performance. Future work includes exploring the potential applications of this algorithm in ISAC systems \cite{chen2023ratesplitting} and initialization-free variants of the proposed algorithm with faster convergence rate.

	\bibliographystyle{IEEEtran}
	\bibliography{reference}	
\end{document}